\documentclass[aps,prl,reprint,superscriptaddress,longbibliography,floatfix]{revtex4-2}
\usepackage{amsmath,amssymb,bm,graphicx,booktabs,microtype,xcolor}

\usepackage[hidelinks]{hyperref}
\newcommand{\ii}{\mathrm{i}}
\newcommand{\dd}{\mathrm{d}}
\DeclareMathOperator{\Tr}{Tr}
\newcommand{\supp}{Supplemental Material}
\newcommand{\tauPl}{\tau_{\rm Pl}}

\begin{document}
\title{Planckian Charge Equilibration in Fast Scramblers}
\author{Jaewon Kim}
\affiliation{Department of Physics, University of Illinois Urbana-Champaign, Urbana, Illinois 61801, USA}
\author{Eduardo Fradkin}
\affiliation{Department of Physics, University of Illinois Urbana-Champaign, Urbana, Illinois 61801, USA}

\begin{abstract}
The Planckian time, $\tauPl=\hbar/(k_BT)$, is conjectured to set a lower limit on thermal relaxation times in strongly interacting quantum matter.
Rigorous bounds limit both how rapidly quantum information can scramble and how soon hydrodynamics can emerge, yet the relation between the scrambling and equilibration rates remains unclear.
We address this question by studying the relaxation of a charge imbalance between two subsystems of the complex Sachdev--Ye--Kitaev model and its Yukawa generalizations, which exhibit maximal scrambling.
In the low-temperature conformal regime, we find through a linear response analysis that the slowest decay of this imbalance occurs at rate $\Gamma=2\pi(2\Delta-d)k_BT/\hbar$, where $\Delta$ is the fermion scaling dimension and $d=0,1$ is the spatial dimension.
These results establish Planckian charge equilibration in maximally scrambling systems and show that maximal scrambling alone does not fix its dimensionless prefactor.
\end{abstract}
\maketitle

Evidence for Planckian relaxation rates of order $k_BT/\hbar$ spans strongly correlated metals with markedly different microscopic interactions and experimental probes ranging from dc transport~\cite{Bruin2013,Legros2019,Licciardello2019,Cao2020,Grissonnanche2021,Ghiotto2021,chang2022} to optical conductivity~\cite{Collins1989,VanDerMarel2003,Prochaska2020} and angle-resolved photoemission~\cite{Valla1999,Reber2019}.
The recurrence of this thermal timescale across diverse materials suggests that the Planckian time, $\tau_{\rm Pl}=\hbar/(k_BT)$, may reflect a fundamental limit on many-body equilibration~\cite{Zaanen2019,HartnollMackenzie2022}.

\begin{figure}[t]
 \centering
 \includegraphics[width=\columnwidth]{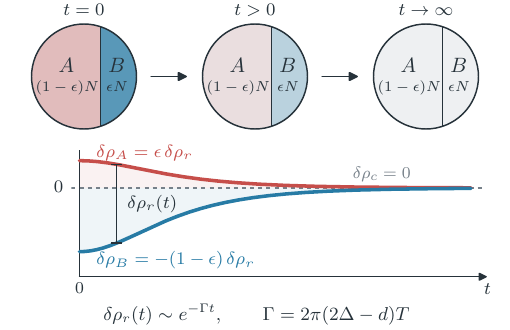}
 \caption{Charge equilibration between subsystems A and B.
 Subsystems A and B contain $N_A=(1-\epsilon)N$ and $N_B=\epsilon N$ fermion flavors, respectively.
 Red and blue indicate excess and deficit density per flavor.
 After a weak chemical-potential bias is removed at $t=0$, the deviations $\delta\rho_A=\epsilon\delta\rho_r$ and $\delta\rho_B=-(1-\epsilon)\delta\rho_r$ relax while their weighted mean remains zero.
 The uniform imbalance $\delta\rho_r=\delta\rho_A-\delta\rho_B$ decays at late times as $e^{-\Gamma t}$, with $\Gamma=2\pi(2\Delta-d)T$ in the conformal regime ($\hbar=k_B=1$).
 }
 \label{fig:equilibration}
\end{figure}

To date, two rigorous Planckian bounds have been demonstrated.
The first concerns quantum scrambling, which spreads initially local quantum information into many-body correlations within an exponentially large Hilbert space~\cite{SekinoSusskind2008,Hosur2016}.
Ref.~\cite{MaldacenaShenkerStanford2016} showed that the quantum Lyapunov exponent $\lambda_L$, which characterizes the rate of scrambling, is bounded by $\lambda_L\le2\pi \tauPl^{-1}$.
The second concerns local equilibration, after which a hydrodynamic description becomes valid.
Ref.~\cite{QiMilekhinDelacretaz2026} established an upper bound $\tau_{\rm hyd}^{-1} \le \alpha \tauPl^{-1}$ on the local equilibration rate defined through a thermally regulated conserved-density correlator, with $\alpha$ determined by the spatial dimension and the type of hydrodynamic behavior.

These two bounds constrain contrasting aspects of many-body dynamics: the spreading of quantum information and the emergence of a hydrodynamic description.
Together, they raise a natural question: how quickly do maximally scrambling systems thermalize and approach the hydrodynamic regime, as probed by the equilibration of charge imbalances?
Furthermore, both Planckian bounds employ thermally regularized correlators, raising a complementary question: does the ordinary retarded density response also exhibit Planckian charge equilibration?

Here we investigate the relation between scrambling and equilibration through charge redistribution in the complex Sachdev--Ye--Kitaev (SYK) model~\cite{SachdevYe1993,MaldacenaStanford2016,Gu2020} and its Yukawa generalizations~\cite{EsterlisSchmalian2019,Bi2017,KimCaoAltman2020,KimCaoAltmanScrambling2020,KimAltmanCao2021,WangChubukov2020,DavisWang2023}, which admit strongly interacting conformal regimes with maximal scrambling at large $N$.
To this end, we divide the fermion flavors into two subsystems $A$ and $B$, and use linear response to study the equilibration of a weak charge imbalance between them, prepared by a quench of a chemical potential imbalance between $A$ and $B$ (Fig.~\ref{fig:equilibration}).
We find that the slowest relative-charge modes in the density response---without thermal regularization---decay with the Planckian rate $\Gamma=2\pi(2\Delta-d)k_BT/\hbar$, where $\Delta$ is the fermion scaling dimension and $d=0,1$ is the spatial dimension.
These results establish Planckian charge equilibration in maximally scrambling models even in ordinary linear response without thermal regularization, while showing that maximal scrambling alone does not fix the rate’s dimensionless prefactor.

\begin{figure}[t]
 \includegraphics[width=\columnwidth]{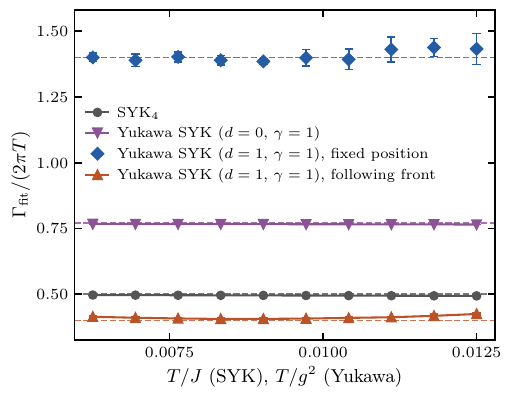}
 \caption{Charge-equilibration rates as a function of temperature for the SYK$_4$ model and its relativistic Yukawa generalizations in $d=0,1$.
 Both Yukawa models have rank ratio $\gamma=1$.
 Temperatures are measured in units of $J$ for SYK$_4$ and $g^2$ for the Yukawa models.
 Dashed lines mark the conformal predictions $\Gamma/(2\pi T)=0.5$, $0.769$, $1.4$, and $0.4$ for SYK$_4$, $d=0$ Yukawa, $d=1$ Yukawa at fixed position, and $d=1$ Yukawa along the ballistic front, respectively.
 Markers and error bars show the midpoint and half-range of decay rates obtained by varying the fitting window.}
 \label{fig:main}
\end{figure}

\emph{Model and Quench.---}
To study the relation between scrambling and equilibration, we consider the SYK model and its Yukawa generalizations, described by the following Euclidean Lagrangian,
\begin{equation}
\begin{split}
\label{eq:H}
\mathcal L &= \sum_{i=1}^{N}\bar\psi_i\bm{\gamma}^\mu\partial_\mu\psi_i - \frac12\sum_{ij,kl}J_{ij,kl}
 :\!(\bar\psi_i\psi_j)(\bar\psi_k\psi_l)\!:\ \\
& \textrm{ where } J_{ij,kl} = \sum_{n=1}^{M} g_{ij}^{n} g_{kl}^{n} \,.
\end{split}
\end{equation}
Here, $\bar\psi=\psi^\dagger\bm{\gamma}^0$, and $\bm{\gamma}^\mu$ are Euclidean gamma matrices satisfying the Clifford algebra.
The fermions $\psi_i$ have one complex component in $0+1$ dimensions, with $\bm{\gamma}^0=1$, and two in $1+1$ dimensions, where we choose $\bm{\gamma}^0=\sigma_x$ and $\bm{\gamma}^1=\sigma_y$.
With the fermion velocity set to one, the kinetic term is then $\psi_i^\dagger(\partial_\tau+\ii\sigma_z\partial_x)\psi_i$, so the two components of $\psi_i$ describe left- and right-moving fermions.
The interaction allows charge transfer between these components, explicitly breaking the continuous chiral $U(1)$ symmetry while preserving total charge.

The interaction matrix $J_{ij,kl}$ is rank $M$ and is constructed from independent, zero-mean Gaussian Hermitian matrices $g^n$, whose elements satisfy
\begin{equation}\label{eq:coupling-variance}
 \overline{g_{ij}^n g_{kl}^m} = g^2 \delta_{il} \delta_{jk} \delta_{nm} / N^2\,.
\end{equation}
The rank ratio $\gamma=M/(n_sN)$ counts the number of Yukawa channels per fermion component, where $n_s$ is the dimension of the Dirac spinor.
In $d=0$, taking $\gamma\to\infty$ recovers the complex SYK$_4$ model~\cite{KimCaoAltman2020}.

These models admit low-temperature conformal regimes in which the fermion scaling dimension $\Delta$ depends continuously on the rank ratio $\gamma$, and exhibit maximal scrambling in $d=0,1$~\cite{KimCaoAltman2020,KimAltmanCao2021}\footnote{The maximal scrambling contribution in $d=1$ propagates near the light cone~\cite{KimAltmanCao2021}.}.
Their zero-temperature equilibrium Green functions take the forms
\begin{subequations}\label{eq:Gc_zeroT}
 \begin{align}
 G_{\mathrm{conf}}(\tau)
 &=-b\,\frac{\operatorname{sgn}\tau}{|\tau|^{2\Delta}},
 \qquad d=0,\\
 G_{\mathrm{conf}}(\tau,x)
 &=-b\,\frac{\tau\bm{\gamma}^0+x\bm{\gamma}^1}
 {(\tau^2+x^2)^{\Delta+1/2}},
 \qquad d=1,
 \end{align}
\end{subequations}
where $b$ is a model-dependent normalization.
In $d=0$, $\Delta\to1/4$ in the SYK$_4$ limit $\gamma\to\infty$.
In $d=1$, this conformal regime is Lorentz invariant and self-tuned for $\gamma>1/4$, and the system flows to an interacting gapless fixed point without tuning a microscopic coupling to a critical value~\cite{KimAltmanCao2021}.

To study equilibration, we partition the fermion flavors into two macroscopic subsystems, $N_A=(1-\epsilon)N$ and $N_B=\epsilon N$, with densities per flavor $\rho_a(x)=N_a^{-1}\sum_{i\in a}:\!\psi_i^\dagger(x)\psi_i(x)\!:$ for $a\in\{A,B\}$, measured relative to the charge-neutral state at zero chemical potential.
We prepare an initial imbalance using density sources $u_a(x)$ coupled as $-\sum_a N_a\int\dd^dx\,u_a(x)\rho_a(x)$.
To distinguish the conserved total-charge channel from the imbalance that relaxes through charge transfer between the subsystems, we define the center (flavor-averaged) and relative densities as
\begin{equation}\label{eq:densities}
 \rho_c=(1-\epsilon)\rho_A+\epsilon\rho_B,
 \qquad \rho_r=\rho_A-\rho_B.
\end{equation}
Introducing the corresponding sources $u_c=(1-\epsilon)u_A+\epsilon u_B$ and $u_r=u_A-u_B$, the source coupling becomes
\begin{equation}\label{eq:source-coupling}
 -N\int\dd^dx\,\bigl[u_c\rho_c+\epsilon(1-\epsilon)u_r\rho_r\bigr].
\end{equation}

To study how the nonconserved relative density relaxes, we prepare a thermal state at temperature $T$ with $u_c=0$ and a weak relative chemical potential $u_r(x)=\mu_r(x)$.
We then switch off the source at $t=0$ and follow the ensuing equilibration.
In $d=0$, the preparation reduces to a single chemical-potential difference $\mu_r$.
In $d=1$, we choose the point source $\mu_r(x)=\mu_r\delta(x)$ to resolve both spatial propagation and charge transfer.

\emph{Bilocal Effective Action.---}
Upon averaging over the disorder and integrating out the fermions, the effective action admits a bilocal description in terms of the Green's functions $G_a(12)=-N_a^{-1}\sum_{i\in a}\langle\mathcal T_\tau\psi_i(1)\bar\psi_i(2)\rangle$ and their self-energies $\Sigma_a$, where $1,2$ denote Euclidean spacetime coordinates.
Defining the averaged Green's function $\overline G=(1-\epsilon)G_A+\epsilon G_B$, the large-$N$ action takes the form (see Appendix.~\ref{SMsec:bilocal})
\begin{equation}\label{eq:bilocal}
 \begin{aligned}
 \frac{1}{N} I[G_a, \Sigma_a]
 ={}&-\sum_{a=A,B}\frac{N_a}{N}
 \Tr\ln(\mathcal G_0^{-1}+u_a\bm{\gamma}^0-\Sigma_a)\\
 &-\sum_{a=A,B}\frac{N_a}{N}
 \Tr(\Sigma_aG_a)+\Phi[\overline G],\\
 \Phi[\overline G]={}&\frac{M}{2N}\Tr\ln(\mathbb I+g^2\mathcal B_{\overline G}).
 \end{aligned}
\end{equation}
We use $\Tr$ for the operator trace over spacetime and any spinor indices present.
Here $\mathcal G_0^{-1}=-\bm{\gamma}^\mu\partial_\mu$ is the bare inverse propagator and $\mathbb I(12)=\delta(1-2)$ is the identity kernel.
Note that the interaction functional $\Phi[\overline G]$ depends only on the averaged Green's function through the fermion bubble $\mathcal B_{\overline G}(12)=\operatorname{tr}[\overline G(12)\overline G(21)]$, with $\operatorname{tr}$ over spinor indices.
Varying the action with respect to $G_A$ and $G_B$ therefore gives identical interaction self-energies, $\Sigma_A=\Sigma_B$, even in the presence of the probe.
The resulting saddle-point equations are
\begin{subequations}\label{eq:saddle}
 \begin{align}
 &G_a^{-1}=\mathcal G_0^{-1}+u_a\bm{\gamma}^0-\Sigma_a,\label{eq:saddle-dyson}\\
 &\Sigma_A(12)=\Sigma_B(12)=\frac{M}{N}g^2\mathcal D_{\overline G}(12)\overline G(12),\label{eq:saddle-self-energy}
 \end{align}
\end{subequations}
where $\mathcal D_{\overline G}=(\mathbb I+g^2\mathcal B_{\overline G})^{-1}$.

To obtain the density susceptibility governing the relaxation of the imbalance, we expand Eq.~\eqref{eq:bilocal} around the source-free equilibrium saddle determined by Eq.~\eqref{eq:saddle}, where $G_A=G_B=G_\star$ and $\Sigma_A=\Sigma_B=\Sigma_\star$ at $u_c=u_r=0$.
We expand the bilocal fields around this saddle as $G_a=G_\star+\delta G_a$ and $\Sigma_a=\Sigma_\star+\delta\Sigma_a$, treating the sources and fluctuations at the same order.
The source-free terms linear in the fluctuations vanish by the saddle-point equations, so the leading response follows from the quadratic action.
We decompose the bilocals into the center fluctuation $\delta G_c=(1-\epsilon)\delta G_A+\epsilon\delta G_B$ and the relative fluctuation $\delta G_r=\delta G_A-\delta G_B$, with $\delta\Sigma_c$ and $\delta\Sigma_r$ defined analogously.
Their equal-time density components give the fluctuations of $\rho_c$ and $\rho_r$.
Since $\overline G=G_\star+\delta G_c$, the interaction functional $\Phi[\overline G]$ depends only on $\delta G_c$ and contributes no relative-channel interaction vertex.
The remaining quadratic terms also separate in this weighted basis, giving $I^{(2)}=I_c^{(2)}+I_r^{(2)}$ with decoupled center and relative channels.
The two actions are
\begin{subequations}\label{eq:quadratic_channels}
 \begin{gather}
 \begin{aligned}
 \frac{I_c^{(2)}}{N}={}&\Tr\!\left[\tfrac12\bigl(G_\star(\delta\Sigma_c-u_c\bm{\gamma}^0)\bigr)^2-\delta\Sigma_c\,\delta G_c\right]\\
 &+\Phi^{(2)}[\delta G_c],
 \end{aligned}\label{eq:Ic}\\
 \frac{I_r^{(2)}}{N\epsilon(1-\epsilon)}=\Tr\!\left[\tfrac12\bigl(G_\star(\delta\Sigma_r-u_r\bm{\gamma}^0)\bigr)^2-\delta\Sigma_r\,\delta G_r\right].\label{eq:Ir}
 \end{gather}
\end{subequations}
For brevity, we omit the explicit expression for the quadratic interaction term $\Phi^{(2)}[\delta G_c]$ here (see Appendix.~\ref{SMsubsec:fluctuations} for its full form).

Since $I_r^{(2)}$ contains no interaction vertex, integrating out its bilocal fluctuations and taking functional derivatives with respect to the relative source $u_r$ gives the relative-density correlator as a bubble:
\begin{equation}\label{eq:bubble}
 \begin{aligned}
 \chi_{rr}^E(12)&=\langle\mathcal T_\tau\rho_r(1)\rho_r(2)\rangle_{\mathrm{conn}}
 =\frac{\Pi(12)}{N\epsilon(1-\epsilon)},\\
 \Pi(12)&=-\operatorname{tr}[\bm{\gamma}^0G_\star(12)\bm{\gamma}^0G_\star(21)] \,.
 \end{aligned}
\end{equation}
Therefore, the leading large-$N$ relative-charge response is determined entirely by the interacting equilibrium fermion propagator.

\emph{Conformal response and the equilibration exponent.---}
At low temperatures, the finite-temperature propagators follow from a conformal mapping of the zero-temperature expressions in Eq.~\eqref{eq:Gc_zeroT}~\cite{Gu2020,KimCaoAltman2020,KimAltmanCao2021}:
\begin{subequations}\label{eq:Gc}
 \begin{align}
 G_{\star,\mathrm{conf}}(\tau)&=-bf(\tau)^{2\Delta},\label{eq:Gc_d0}\\
 G_{\star,\mathrm{conf}}(\tau,x)&=-b(f_+f_-)^{\Delta-1/2}
 \begin{pmatrix}0&f_+\\f_-&0\end{pmatrix},\label{eq:Gc_d1}
 \end{align}
\end{subequations}
where $f(\tau)=\pi T/\sin(\pi T\tau)$ and $f_\pm=f(\tau\pm\ii x)$, with $0<\tau<\beta$.
The positive Euclidean product $f_+f_-$ fixes the fractional-power branch, while the matrix factor ensures fermionic antiperiodicity.

Substituting Eq.~\eqref{eq:Gc} into Eq.~\eqref{eq:bubble} and analytically continuing to real time gives the retarded correlator $\chi_{rr}^R(x,t)=\ii\theta(t)\langle[\rho_r(x,t),\rho_r(0,0)]\rangle$.
Writing $F(t)=\pi T/\sinh(\pi Tt)$ and $F_\pm=F(t\pm x)$, the $d=0$ and $d=1$ responses are, respectively,
\begin{subequations}\label{eq:retarded_response}
 \begin{align}
 \chi_{rr}^R(t)&=\frac{\mathcal C_\Delta}{N}\theta(t)F(t)^{4\Delta},\label{eq:retarded_d0}\\
 \chi_{rr}^R(x,t)&=\frac{\mathcal C_\Delta}{N}\theta(t-|x|)(F_+F_-)^{2\Delta-1}(F_+^2+F_-^2).\label{eq:retarded_d1}
 \end{align}
\end{subequations}
Here $\theta$ is the Heaviside function and $\mathcal C_\Delta=2b^2\sin(2\pi\Delta)/[\epsilon(1-\epsilon)]$.
In $d=1$, the response vanishes outside the light cone, $|x|>t>0$, as required by causality.
Note that Eq.~\eqref{eq:retarded_response} already displays exponential decay of the relative-density response:
for $Ty\gg1$, $F(y)\sim2\pi T e^{-\pi Ty}$.

To be more concrete, we compute the density imbalance remaining at time $t>0$ after switching off the source at $t=0$.
In the linear-response regime, the remaining imbalance is given by a time integral of the retarded correlator:
\begin{equation}\label{eq:quench}
 \delta\rho_r(x,t)=N\epsilon(1-\epsilon)\mu_r\int_t^\infty\dd u\,\chi_{rr}^R(x,u).
\end{equation}
In $d=0$, the spatial argument is omitted.

For $d=0$, the response at $Tt\gg1$ is
\begin{equation}\label{eq:quench_d0}
 \delta\rho_r(t)\simeq \mu_r P_\Delta e^{-4\pi \Delta T t},
\end{equation}
where $P_\Delta=b^2\sin(2\pi\Delta)(2\pi T)^{4\Delta-1}/\Delta$.
Consequently, the subsystem density imbalance decays at a Planckian rate of $4\pi\Delta T$.

For $d=1$, the prepared density profile remains unchanged until the ballistic front arrives at $t=|x|$.
At fixed position $x$ well behind the ballistic front, $T(t-|x|)\gg1$, the imbalance decays at rate $4\pi\Delta T$:
\begin{equation}\label{eq:quench_d1}
 \delta\rho_r(x,t)\simeq2\mu_r P_\Delta\cosh(2\pi Tx)e^{-4\pi\Delta Tt}.
\end{equation}

On the other hand, the imbalance along the ballistic front decays at the slower rate $2\pi(2\Delta-1)T$.
At fixed distance $s>0$ behind the front and far from the initial source, $T(t-s)\gg1$, the response is
\begin{equation}\label{eq:quench_d1_front}
 \delta\rho_r\bigl(\pm(t-s),t\bigr)\simeq\mu_r P_\Delta
 \frac{e^{-2\pi(2\Delta-1)T(t-s)}}{\bigl(e^{2\pi Ts}-1\bigr)^{2\Delta}}.
\end{equation}
Note that $P_\Delta<0$ in $d=1$, so the density response $\delta\rho_r/\mu_r$ is negative behind the ballistic front.
This negative tail is also observed in our low-temperature numerics (see Fig.\ref{fig:decay-fits}).

By translation invariance, spatial integration of the full retarded kernel in Eq.~\eqref{eq:quench} gives the response at any point to a uniform quench $\mu_r(x)=\mu$.
The resulting density imbalance also decays at the slower rate $2\pi(2\Delta-1)T$ (see Sec.~2.2 of the \supp).

Consequently, the slowest relative-density decay governing subsystem equilibration has decay rate
\begin{equation}\label{eq:Gamma}
    \Gamma=2\pi(2\Delta-d)T
\end{equation}
in $d=0,1$.
This indicates that charge equilibration between subsystems in these maximally scrambling models is Planckian, with a nonuniversal $O(1)$ prefactor set by the fermion scaling dimension $\Delta$ \footnote{This characterization concerns the interior of the conformal regime. The endpoint limits $\Delta\to0,1/2$ in $d=0$ and $\Delta\to1/2,1$ in $d=1$ are discussed in Sec.~2 of the \supp.}.

In the families of maximally scrambling models studied here, the fermion conformal scaling dimension satisfies $0<\Delta<1/2$ in $d=0$ and $1/2<\Delta<1$ in $d=1$~\cite{KimCaoAltman2020,KimAltmanCao2021} \footnote{The positive-semidefinite Yukawa ensemble in Eq.~\eqref{eq:H} realizes $1/4\le\Delta<1/2$ in $d=0$; the complementary range $0<\Delta<1/4$ belongs to the negative-semidefinite ensemble~\cite{KimCaoAltman2020}.}.
For the slowest relative-density mode, Eq.~\eqref{eq:Gamma} then implies $0<\Gamma<2\pi T$.
The slowest relative-charge equilibration rate therefore lies below the same $2\pi T$ ceiling that bounds the quantum Lyapunov exponent of Ref.~\cite{MaldacenaShenkerStanford2016}.

\emph{Numerical verification.---}
We test the conformal prediction by numerically computing the relaxation of a weak charge imbalance in complex SYK$_4$ and its Yukawa generalizations in $d=0,1$.
We evaluate the relative-density susceptibility from the full equilibrium real-time Green's functions and use linear response to compute the time evolution of the density imbalance.
Fig.~\ref{fig:main} shows the rate estimates approaching $\pi T$ for SYK$_4$ and $4\pi\Delta T$ for the $d=0$ Yukawa model, where $\Delta\simeq0.3847$ at $\gamma=1$.
For the $d=1$ Yukawa model~\footnote{The $d=1$ theory arises as the low-energy limit of a half-filled, translation-invariant fermion chain with nearest-neighbor hopping and on-site all-to-all flavor interactions~\cite{KimAltmanCao2021}; see Sec.~3.3 of the \supp{} for the microscopic Hamiltonian and numerical parameters.}, the rates likewise approach $4\pi\Delta T$ at fixed position and $2\pi(2\Delta-1)T$ along the moving front, with $\Delta=0.7$ at $\gamma=1$.

\emph{Discussion.---}
Our results establish Planckian charge equilibration in the SYK model and its Yukawa generalizations.
Although the models studied here share the maximal scrambling rate $\lambda_L=2\pi T$, their relative charge-equilibration rate $\Gamma=2\pi(2\Delta-d)T$ retains a nonuniversal prefactor determined by the fermion scaling dimension $\Delta$.
Nevertheless, throughout these conformal regimes, $\Gamma<2\pi T$, giving the slowest relative-density mode the same ceiling as the quantum Lyapunov exponent.
Whether this shared ceiling reflects a common constraint on scrambling and equilibration remains an open question.
Furthermore, whereas the rigorous bounds on scrambling and hydrodynamic onset employ thermally regularized correlators~\cite{MaldacenaShenkerStanford2016,QiMilekhinDelacretaz2026}, the Planckian timescale found here emerges directly in the ordinary retarded density response.

The relative-charge response studied here determines the decay of the nonconserved contribution to the subsystem density response.
In $d=1$, establishing the approach to charge hydrodynamics also requires understanding the center response, which governs conserved-charge transport.
We conjecture that this channel crosses over from ballistic propagation at short times to diffusion at long times as the charge current relaxes.
We leave the characterization of this crossover and its timescale for future work.

While the subsystems studied here are distinguished by flavor, an intriguing possibility in a metal is that they could instead comprise different regions of the Fermi surface.
If nonconserved imbalances between momentum sectors also relax on a Planckian timescale, rapid redistribution could contribute to incoherent electronic dynamics.
Understanding how this mixing enters optical and transport responses would connect subsystem equilibration to experimentally measured relaxation in quantum critical metals.

\emph{Acknowledgments.---}
We would like to thank Xiangyu Cao, Shubhayu Chatterjee, Jongyeon Lee, Aleksey Milekhin, and Marvin Qi for helpful discussions.
Jaewon Kim acknowledges support from the National Science Foundation, Grant No. DMR-2225920.
Part of this work was done in the Aspen Center for Physics, supported by the National Science Foundation, Grant No. PHY-2210452.

\bibliography{references}

\begin{widetext}

\appendix
\setcounter{secnumdepth}{3}

\section{Bilocal Action}
\label{SMsec:bilocal}

In this section, we derive the large-$N$ bilocal action Eq.\eqref{eq:bilocal} of the main text that describes the dynamics of the complex SYK model and its Yukawa generalizations.
We then derive its fluctutation expansion Eq.\eqref{eq:quadratic_channels} to the relative and center channels, which in turn determine the relative and center density response.

Upon a Hubbard--Stratonovich (HS) transformation, we can rewrite the Lagrangian of Eq.~\eqref{eq:H} in the main text in terms of $M$ real HS fields interacting with the Dirac fermions through a Yukawa interaction, including the subsystem chemical potentials $u_a$:
\begin{equation}\label{eq:yukawa-lagrangian}
 \mathcal L=\sum_{a=A,B}\sum_{i\in a}\bar\psi_i(\bm{\gamma}^\mu\partial_\mu-u_a\bm{\gamma}^0)\psi_i
 +\frac12\sum_{n=1}^{M}\phi_n^2
 +\sum_{i,j=1}^{N}\sum_{n=1}^{M}g_{ij}^{n}\phi_n\bar\psi_i\psi_j.
\end{equation}

We perform the quenched disorder average at the replica-diagonal large-$N$ saddle.
Denoting the Yukawa contribution to the Euclidean action by $I_{\rm int}$, averaging over the Gaussian matrices $g^n$ gives
\begin{equation}\label{eq:yukawa-disorder-average}
 \overline{e^{-I_{\rm int}}}
 =\exp\!\left[\frac{g^2}{2N^2}\sum_n\int_{12}\phi_n(1)\phi_n(2)
 \sum_{ij}\bigl[\bar\psi_i(1)\psi_j(1)\bigr]
                 \bigl[\bar\psi_j(2)\psi_i(2)\bigr]\right].
\end{equation}
Here $1=(\tau_1,x_1)$ and $\int_{12}=\int_1\int_2$, with $\int_1=\int_0^\beta\dd\tau_1\int\dd^d x_1$ and spatial coordinates omitted in $d=0$.

We now introduce independent bilocal fields constrained to equal the fermion bilinears,
\begin{equation}\label{eq:bilocal-constraint}
 G_a(12)=-\frac{1}{N_a}\sum_{i\in a}\psi_i(1)\bar\psi_i(2),
 \qquad
 \overline G=\sum_a f_aG_a,
 \qquad f_a=\frac{N_a}{N}=\begin{cases}
 1-\epsilon,& a=A,\\
 \epsilon,& a=B.
 \end{cases}
\end{equation}
To enforce Eq.~\eqref{eq:bilocal-constraint}, we introduce bilocal Lagrange multipliers $\Sigma_a$ through the functional identity
\[
 1\propto\int\prod_a\mathcal D G_a\,\mathcal D\Sigma_a\,
 \exp\!\left\{\sum_a N_a\Tr\!\left[
 \Sigma_a\left(G_a+\frac{1}{N_a}\sum_{i\in a}\psi_i\bar\psi_i\right)\right]\right\}.
\]
Here, $\Tr$ denotes the operator trace over spacetime and any spinor indices present.
At the saddle, $G_a$ is the Green function $-N_a^{-1}\sum_{i\in a}\langle\mathcal T_\tau\psi_i(1)\bar\psi_i(2)\rangle$.
It is a scalar in $d=0$ and a $2\times2$ spinor matrix in $d=1$.

Reordering the Grassmann fields in Eq.~\eqref{eq:yukawa-disorder-average} gives
\[
 \sum_{ij}\bigl[\bar\psi_i(1)\psi_j(1)\bigr]
             \bigl[\bar\psi_j(2)\psi_i(2)\bigr]
 =-N^2\operatorname{tr}_{\rm spin}[\overline G(12)\overline G(21)].
\]
Thus the disorder-averaged interaction depends only on the flavor average $\overline G$.

At fixed $G_a$ and $\Sigma_a$, the fermionic and bosonic integrals are Gaussian.
Performing these integrals gives the bilocal action of Eq.~\eqref{eq:bilocal} in the main text, up to field-independent constants:
\begin{equation}\label{eq:action}
 \begin{aligned}
 \frac{I[G_a,\Sigma_a]}{N}
 &=-\sum_a f_a\Tr\ln(\mathcal G_0^{-1}+u_a\bm{\gamma}^0-\Sigma_a)
 -\sum_a f_a\Tr(\Sigma_aG_a)+\Phi[\overline G],\\
 \Phi[\overline G]&=\frac{M}{2N}\Tr\ln(\mathbb I+g^2\mathcal B_{\overline G}),\\
 \mathcal B_{\overline G}(12)&=\operatorname{tr}_{\rm spin}[\overline G(12)\overline G(21)].
 \end{aligned}
\end{equation}
The two logarithms arise from the fermion and HS-boson Gaussian integrals, respectively, while $-\sum_a f_a\Tr(\Sigma_aG_a)$ comes from the bilocal constraint.
Here $\mathcal G_0^{-1}=-\bm{\gamma}^\mu\partial_\mu$ is the bare inverse fermion propagator and $\mathbb I(12)=\delta(1-2)$ is the identity kernel.
The trace in the Luttinger-Ward functional $\Phi$ acts on spacetime only, because the spinor trace has already been taken in $\mathcal B_{\overline G}$.
The source kernel is $u_a(1)\bm{\gamma}^0\delta(1-2)$; in $d=0$, $\bm{\gamma}^0=1$ and $\mathcal G_0^{-1}(\ii\nu_n)=\ii\nu_n$.

Varying with regards to $G_a, \Sigma_a$ gives the Schwinger-Dyson equations.
Writing $\mathcal D_{\overline G}=(\mathbb I+g^2\mathcal B_{\overline G})^{-1}$, they are given as,
\begin{equation}\label{eq:yukawa-saddle}
 G_a^{-1}=\mathcal G_0^{-1}+u_a\bm{\gamma}^0-\Sigma_a,
 \qquad
 \Sigma_A(12)=\Sigma_B(12)=\frac{M}{N}g^2\mathcal D_{\overline G}(12)\overline G(12).
\end{equation}
Crucially, note that the self-energies of the subsystems are equal.
This equality follows from the fact that the interaction depends on $G_A,G_B$ only through their weighted average $\overline G$.
This equivlaence is the property that removes the relative-channel interaction vertex in the fluctuation analysis.

\paragraph{The SYK$_4$ limit.}
In $d=0$, the coupling coefficients to the four-fermion interaction $J_{ij;kl}$ have variance $\overline{|J_{ij;kl}|^2}=Mg^4/N^4$ for four distinct indices.
To keep the SYK interaction energy extensive as $M/N\to\infty$, we rescale $g$ so that $J^2=(M/N)g^4$ remains fixed:
\begin{equation}\label{eq:syk-variance}
 g^2=J\sqrt{\frac{N}{M}},
 \qquad
 \overline{|J_{ij;kl}|^2}=\frac{J^2}{N^3}.
\end{equation}
With this rescaling, $g^2$ vanishes as $M/N\to\infty$, so we can expand the Luttinger-Ward functional in powers of $g^2$:
\begin{equation}\label{eq:high-rank-expansion}
 \Phi[\overline G]=\frac{Mg^2}{2N}\Tr\mathcal B_{\overline G}
 -\frac{J^2}{4}\int_{12}\overline G(12)^2\overline G(21)^2.
\end{equation}
Note that higher-order terms vanish as $M/N\to\infty$ at fixed $J$~\cite{KimCaoAltman2020}.

To evaluate the first term, we use the equal-time limits of the $d=0$ equilibrium saddle, $\overline G(\tau,\tau^+)=\tfrac12+\rho_c$ and $\overline G(\tau^+,\tau)=\rho_c-\tfrac12$, where $\rho_c$ is the equilibrium density per flavor relative to half filling and $\tau^+=\tau+0^+$.
With this point-splitting prescription,
\begin{equation}\label{eq:trace-b}
 \begin{aligned}
 \Tr\mathcal B_{\overline G}
 &=\int_0^\beta\dd\tau\,\overline G(\tau,\tau^+)\overline G(\tau^+,\tau)\\
 &=\int_0^\beta\dd\tau\,\left(\rho_c+\frac12\right)\left(\rho_c-\frac12\right)
 =\beta\left(\rho_c^2-\frac14\right).
 \end{aligned}
\end{equation}
Consequently, the first term of Eq.~\eqref{eq:high-rank-expansion} is fixed by the conserved total charge and does not affect the dynamics.
After subtracting this background, the second term of Eq.~\eqref{eq:high-rank-expansion} gives the complex-SYK$_4$ Luttinger-Ward functional and self-energy~\cite{Gu2020}:
\begin{equation}\label{eq:syk-functional}
 \Phi_{\rm SYK}[\overline G]= -\frac{J^2}{4}\int_{12}\overline G(12)^2\overline G(21)^2,
 \qquad \Sigma_a(12)=-J^2\overline G(12)^2\overline G(21).
\end{equation}
Thus, in the limit $M/N\to\infty$, the $d=0$ model reduces to complex SYK$_4$ within each fixed-charge sector.

\subsection{Quadratic Fluctuation Expansion}
\label{SMsubsec:fluctuations}
To obtain the density susceptibilities, we expand Eq.~\eqref{eq:action} to quadratic order about equilibrium and separate the center and relative channels.
At zero source, the flavor-symmetric saddle has $G_A=G_B=G_\star$ and $\Sigma_A=\Sigma_B=\Sigma_\star$.
We parameterize the bilocal fields as $G_a=G_\star+\delta G_a$ and $\Sigma_a=\Sigma_\star+\delta\Sigma_a$, treating the sources $u_a$ and the fluctuations at the same perturbative order.
The center and relative combinations are $x_c=(1-\epsilon)x_A+\epsilon x_B$ and $x_r=x_A-x_B$, respectively, for $x=\delta G,\delta\Sigma,u$.
The saddle-point equations eliminate the source-free linear terms, giving,
\begin{equation}\label{eq:quadratic}
 \frac{I^{(2)}}{N}
 =\frac12\sum_a f_a\Tr\!\left[G_\star(\delta\Sigma_a-u_a\bm{\gamma}^0)G_\star(\delta\Sigma_a-u_a\bm{\gamma}^0)\right]
 -\sum_a f_a\Tr(\delta\Sigma_a\,\delta G_a)+\Phi^{(2)}[\delta G_c].
\end{equation}
To expand the Luttinger-Ward functional explicitly, write $\mathcal B_{G_\star+\delta G_c}=\mathcal B_\star+\mathcal B_1+\mathcal B_2$, where
\begin{equation}
 \begin{aligned}
 \mathcal B_1(12)&=\operatorname{tr}_{\rm spin}\!\left[G_\star(12)\delta G_c(21)+\delta G_c(12)G_\star(21)\right],\\
 \mathcal B_2(12)&=\operatorname{tr}_{\rm spin}\!\left[\delta G_c(12)\delta G_c(21)\right].
 \end{aligned}
\end{equation}
To quadratic order in the bilocal fluctuations, the Luttinger-Ward functional contributes
\begin{equation}\label{eq:yukawa-quadratic}
 \Phi^{(2)}[\delta G_c]=\frac{M}{2N}
 \left[g^2\Tr(\mathcal D_\star\mathcal B_2)
 -\frac{g^4}{2}\Tr(\mathcal D_\star\mathcal B_1\mathcal D_\star\mathcal B_1)\right].
\end{equation}
These traces act on spacetime kernels, and both terms depend only on $\delta G_c$.
Thus the interaction contribution to the Hessian has no relative-channel component.


Consequently, reparametrizing Eq.~\eqref{eq:quadratic} in terms of the center and relative degrees of freedom, the quadratic fluctuations is given by,
\begin{subequations}\label{eq:quadratic-channels}
 \begin{align}
 I^{(2)}&=I_c^{(2)}[\delta G_c,\delta\Sigma_c,u_c]+I_r^{(2)}[\delta G_r,\delta\Sigma_r,u_r],\notag\\
 \frac1N & I_c^{(2)}[\delta G_c,\delta\Sigma_c,u_c]
 =\Tr\!\left[\tfrac12\bigl(G_\star(\delta\Sigma_c-u_c\bm{\gamma}^0)\bigr)^2-\delta\Sigma_c\,\delta G_c\right]+\Phi^{(2)}[\delta G_c].\label{eq:common-quadratic} \\
 \frac{1}{N\epsilon(1-\epsilon)} & I_r^{(2)}[\delta G_r,\delta\Sigma_r,u_r]
 =\Tr\!\left[\tfrac12\bigl(G_\star(\delta\Sigma_r-u_r\bm{\gamma}^0)\bigr)^2-\delta\Sigma_r\,\delta G_r\right],\label{eq:relative}
 \end{align}
\end{subequations}

Since the center and relative fluctuations decouple in Eq.~\eqref{eq:quadratic-channels}, the partition function factorizes at quadratic order, up to a source-independent normalization:
\begin{equation}\label{eq:channel-partition-functions}
 \begin{aligned}
 Z[u_c,u_r]&\propto Z_c[u_c]Z_r[u_r],\\
 Z_c[u_c]&=\int\mathcal D(\delta G_c)\,\mathcal D(\delta\Sigma_c)\,e^{-I_c^{(2)}[\delta G_c,\delta\Sigma_c,u_c]},\\
 Z_r[u_r]&=\int\mathcal D(\delta G_r)\,\mathcal D(\delta\Sigma_r)\,e^{-I_r^{(2)}[\delta G_r,\delta\Sigma_r,u_r]}.
 \end{aligned}
\end{equation}
Here $Z_c$ and $Z_r$ are the partition functions of the center and relative fluctuations, respectively, with sources coupled to density deviations from half filling.
The source contribution to the Euclidean action, with $1=(\tau_1,x_1)$, is
\[
 -\sum_a N_a\int_1 u_a(1)\rho_a(1)
 =-N\int_1\left[u_c(1)\rho_c(1)+\epsilon(1-\epsilon)u_r(1)\rho_r(1)\right].
\]
Accordingly, the connected density correlators follow from the functional derivatives
\begin{subequations}\label{eq:chi-derivative}
 \begin{align}
 \chi_{cc}^E(12)&=\langle\mathcal T_\tau\rho_c(1)\rho_c(2)\rangle_{\rm conn}
 =\frac{1}{N^2}\left.\frac{\delta^2\ln Z_c[u_c]}{\delta u_c(1)\delta u_c(2)}\right|_{u_c=0},\\
 \chi_{rr}^E(12)&=\langle\mathcal T_\tau\rho_r(1)\rho_r(2)\rangle_{\rm conn}
 =\frac{1}{[N\epsilon(1-\epsilon)]^2}\left.\frac{\delta^2\ln Z_r[u_r]}{\delta u_r(1)\delta u_r(2)}\right|_{u_r=0}.
 \end{align}
\end{subequations}

\section{Relative Channel Response}
\label{SMsec:relative-channel}
\label{sec:conformal-continuation}
To obtain the relative-density response, we integrate out $\delta G_r$ and $\delta\Sigma_r$ of Eq.\eqref{eq:channel-partition-functions}.
As the action is quadratic, evaluating it at the stationary values of $\delta G_r$ and $\delta\Sigma_r$ gives $Z_r[u_r]\propto e^{-I_{r,*}[u_r]}$ up to a source-independent factor.
The absence of an interaction term makes these stationary values straightforward to determine: varying with respect to $\delta G_r$ and $\delta\Sigma_r$, respectively, gives
\[
 \delta\Sigma_r^*=0,\qquad
 \delta G_r^*=-G_\star u_r\bm{\gamma}^0G_\star.
\]
Substituting these stationary values into the action gives
\begin{equation}
 \frac{1}{N\epsilon(1-\epsilon)}I_{r,*}[u_r]
 =\Tr\!\left[\tfrac12\bigl(G_\star u_r\bm{\gamma}^0\bigr)^2\right]
 =-\frac12\int_{12}u_r(1)\Pi(12)u_r(2).
\end{equation}
Taking the functional derivatives with regards to $u_r$, we find that the relative density correlator is given by,
\begin{equation}\label{appeq:bubble}
 \begin{aligned}
 \chi_{rr}^E(12)&=\langle\mathcal T_\tau\rho_r(1)\rho_r(2)\rangle_{\rm conn}
 =\frac{\Pi(12)}{N\epsilon(1-\epsilon)},\\
 \Pi(12)&=-\operatorname{tr}_{\rm spin}\!\left[\bm{\gamma}^0G_\star(12)\bm{\gamma}^0G_\star(21)\right].
 \end{aligned}
\end{equation}

Or alternatively, for a translation-invariant saddle,
\begin{equation}\label{eq:bubble-momentum}
 \Pi(\ii\Omega_m,q)=-T\sum_n\int\frac{\dd^d k}{(2\pi)^d}
 \operatorname{tr}_{\rm spin}\!\left[
 \bm{\gamma}^0G_\star(\ii\nu_n+\ii\Omega_m,k+q)
 \bm{\gamma}^0G_\star(\ii\nu_n,k)\right],
\end{equation}
with the momentum integral omitted in $d=0$.

We now evaluate the relative-density bubble using conformal propagators and obtain the late-time density imbalance after the quench by linear response.
We first treat $d=0$, then derive the Dirac response in $d=1$.
In $d=1$, we distinguish the signal at a fixed position from the damping of a propagating front and a uniform imbalance.
We also examine the endpoint limits $\Delta\to0,1/2$ in $d=0$ and $\Delta\to1/2,1$ in $d=1$, where the response amplitude must be considered together with its decay rate.

\subsection{General conformal response in zero spatial dimensions}
For a half-filled conformal saddle with fermion dimension $0<\Delta<1/2$, the propagator and relative bubble at $0<\tau<\beta$ are
\begin{equation}
 G_{\star,\mathrm{conf}}(\tau)=-b\left[\frac{\pi T}{\sin(\pi T\tau)}\right]^{2\Delta},
 \qquad
 \Pi^E(\tau)=b^2\left[\frac{\pi T}{\sin(\pi T\tau)}\right]^{4\Delta}.
\end{equation}

With $\Pi^R=N\epsilon(1-\epsilon)\chi_{rr}^R$ and $\chi_{rr}^R(t)=\ii\theta(t)\langle[\rho_r(t),\rho_r(0)]\rangle$, analytic continuation to the two sides of the thermal strip gives
\begin{equation}\label{eq:general-conformal-retarded}
 \Pi^R(t)=\ii\theta(t)\lim_{\eta\downarrow0}
 [\Pi^E(\eta+\ii t)-\Pi^E(\beta-\eta+\ii t)]
 =2b^2\sin(2\pi\Delta)\left[\frac{\pi T}{\sinh(\pi Tt)}\right]^{4\Delta},\qquad t>0.
\end{equation}

For a field present before $t=0$ and removed afterward, linear response gives
\begin{equation}\label{eq:general-conformal-step}
 \begin{aligned}
 \frac{\delta\rho_r(t)}{\mu_r}
 &=\int_t^\infty\dd u\,\Pi^R(u)\\
 &=2b^2\sin(2\pi\Delta)(2\pi T)^{4\Delta-1}
 B_{e^{-2\pi Tt}}(2\Delta,1-4\Delta)\\
 &\underset{Tt\gg1}{\sim}
 \frac{b^2\sin(2\pi\Delta)}{\Delta}(2\pi T)^{4\Delta-1}
 e^{-4\pi\Delta Tt}.
 \end{aligned}
\end{equation}
Here $B_z(a,b)=\int_0^z\dd y\,y^{a-1}(1-y)^{b-1}$ is finite for every $t>0$, even when $1-4\Delta\le0$, because $z=e^{-2\pi Tt}<1$.
The conformal late-time exponent is therefore $\Gamma=4\pi\Delta T$.

\subsubsection{Endpoint limits}
\paragraph{Lower End Point:}
$\Delta \rightarrow 0$.
In this limit, with $b$ approaching a finite value $b_0$, the late-time response in Eq.~\eqref{eq:general-conformal-step} becomes
\begin{equation}\label{eq:small-delta-endpoint}
 \frac{\delta\rho_r(t)}{\mu_r}\sim\frac{b_0^2}{T}e^{-4\pi\Delta Tt},
 \qquad \Gamma=4\pi\Delta T\to0.
\end{equation}
The vanishing decay rate is consistent with the corresponding free endpoint of the Yukawa--SYK model, where the self-energy vanishes and the charges of $A$ and $B$ are separately conserved, so the prepared imbalance persists~\cite{KimCaoAltman2020}.

\paragraph{Upper End Point:}
$\Delta \rightarrow 1/2$.
Writing $\Delta=1/2-\zeta$, with $\zeta>0$, gives $\sin(2\pi\Delta)=2\pi\zeta+O(\zeta^3)$.
For a finite limiting normalization $b$, the amplitudes in Eqs.~\eqref{eq:general-conformal-retarded} and \eqref{eq:general-conformal-step} therefore vanish linearly in $\zeta$, even though $\Gamma\to2\pi T$.
The conformal tail disappears at the endpoint, so its limiting exponent does not determine the microscopic relaxation.
To resolve that relaxation, we must retain the finite bandwidth omitted from the conformal propagator.

At this endpoint, the dynamics are again effectively those of free fermions.
Unlike the $\Delta\to0$ limit, where the effective Hamiltonian vanishes, a nonzero random quadratic Hamiltonian now mixes A and B, allowing an initial imbalance to relax through dephasing.
Both the complex SYK$_2$ model and the condensed subextensive-rank saddle of the Yukawa--SYK model at $M/N\to0$ realize this endpoint and have a semicircular single-particle spectrum~\cite{KimCaoAltman2020}.

In these models, the exact retarded propagator is given by,
\begin{equation}\label{eq:free-quadratic-retarded}
 G_\star^R(t)=-\ii\theta(t)\frac{J_1(2Kt)}{Kt},
\end{equation}
where $J_1$ is the Bessel function of the first kind and $K$ denotes the bandwidth.
At late times, the retarded propagator has an oscillatory envelope that decays as $t^{-3/2}$.
Since the relative-density response is a particle--hole bubble constructed from this finite-band spectrum, we likewise expect the density imbalance to have an algebraic late-time tail, rather than exponential relaxation with a finite thermal decay rate.

\subsection{General conformal response in one spatial dimension}
\label{sec:dirac-conformal-response}

In $d=1$, charge redistribution accompanies spatial propagation.
We derive the response in both spacetime and momentum space to identify the slowest decay rate and separate it from the decay observed at a fixed position.
We set the emergent velocity to one and use $\bm{\gamma}^0=\sigma_x$, $\bm{\gamma}^1=\sigma_y$, and $G_{\star,\mathrm{conf}}=-\langle T_\tau\psi\bar\psi\rangle$, with $\bar\psi=\psi^\dagger\bm{\gamma}^0$.
For $1/2<\Delta<1$, defining
\begin{equation}
 a_\pm=\Delta\pm\frac12,\qquad
 f_\pm(\tau,x)=\frac{\pi T}{\sin[\pi T(\tau\pm\ii x)]},
\end{equation}
the finite-temperature Dirac propagator are given by~\cite{KimAltmanCao2021},
\begin{equation}
 G_{\star,\mathrm{conf}}(\tau,x)
 =-b(f_+f_-)^{\Delta-1/2}
 \begin{pmatrix}0&f_+\\ f_-&0\end{pmatrix}.
 \label{eq:dirac-thermal-propagator}
\end{equation}

Accordingly, the relative-density propagator is given proportional to the bubble,
\begin{align}
 \Pi^E(\tau,x)
 &=-\operatorname{tr}\!\left[
 \bm{\gamma}^0G_{\star,\mathrm{conf}}(\tau,x)\bm{\gamma}^0G_{\star,\mathrm{conf}}(-\tau,-x)\right]\nonumber\\
 &=b^2\left[f_+^{2a_+}f_-^{2a_-}
              +f_+^{2a_-}f_-^{2a_+}\right].
 \label{eq:dirac-thermal-density-bubble}
\end{align}
Here $a_+=\Delta+\tfrac12$ and $a_-=\Delta-\tfrac12$.

To obtain the retarded response, we analytically continue the Euclidean bubble to real time,
\begin{equation}
 \Pi^R(t,x)=\ii\theta(t)\lim_{\eta\downarrow0}
 \left[\Pi^E(\eta+\ii t,x)
       -\Pi^E(\beta-\eta+\ii t,x)\right].
 \label{eq:dirac-retarded-prescription}
\end{equation}
Away from the light cone this is
\begin{align}
 \Pi^R(t,x)&=2b^2\sin(2\pi\Delta)\theta(u)\theta(v)
 \left[F(v)^{2a_+}F(u)^{2a_-}
       +F(v)^{2a_-}F(u)^{2a_+}\right],\nonumber\\[-1mm]
 &\hspace{12mm}u=t-x,\quad v=t+x,\quad
 F(y)=\frac{\pi T}{\sinh(\pi Ty)}.
 \label{eq:dirac-retarded-interior}
\end{align}
Note that Eq.~\eqref{eq:dirac-retarded-interior} applies when both $u$ and $v$ are long compared with microscopic time scales.

Fourier transforming Eq.\eqref{eq:dirac-retarded-interior}, we find,
\begin{equation}
 \Pi^R(\omega,k)=b^2\sin(2\pi\Delta)(2\pi T)^{4\Delta-2}
 \sum_{\sigma=\pm}B_\sigma(\omega-k)B_{-\sigma}(\omega+k),
 \label{eq:dirac-retarded-beta}
\end{equation}
Here $B_\sigma(q)=B(a_\sigma-\ii q/(4\pi T),1-2a_\sigma)$ where $B$ are beta functions, given by $B(z,w)=\Gamma(z)\Gamma(w)/\Gamma(z+w)$.

For real $k$, the poles closest to the real frequency axis come from the beta functions with first argument $a_- -\ii(\omega\pm k)/(4\pi T)$:
\begin{equation}\label{eq:dirac-slowest-poles}
 \omega_\pm(k)=\pm k-\ii\Gamma,
 \qquad \Gamma=2\pi T(2\Delta-1).
\end{equation}
Consequently, for all momenta in the conformal regime, $\Pi^R(t,k)$ has the late-time envelope $e^{-\Gamma t}$, with oscillations at frequency $|k|$ for $k\ne0$.

At $k=0$, which probes a spatially uniform imbalance, the two contributions add to give a simple pole:
\begin{equation}
 \Pi^R(\omega,0)\underset{\omega\to-\ii\Gamma}{\sim}
 \frac{\mathcal A_\Delta}{\Gamma-\ii\omega},\qquad
 \mathcal A_\Delta=-\frac{2b^2\sin(2\pi\Delta)}{\Delta}(2\pi T)^{4\Delta-1}>0.
 \label{eq:dirac-integrated-relaxation-pole}
\end{equation}
For a spatially uniform source $\mu_r$ switched off at $t=0$, the late-time density imbalance is
\begin{equation}\label{eq:dirac-field-removal-tail}
 \delta\rho_r(t)\sim\mu_r\frac{\mathcal A_\Delta}{\Gamma}e^{-\Gamma t}.
\end{equation}

\subsubsection{Spatially localized field removal}
We now apply the retarded kernel to a localized preparation, computing both the late-time signal at a fixed observation point, and that of the envelope traveling near the light cone.

For a weak source $\mu_r(x)$ present before $t=0$ and removed afterward, translational invariance gives
\begin{equation}\label{eq:local_step}
 \delta\rho_r(x,t)=\int\dd x'\,\mu_r(x')\int_t^\infty\dd u\,\Pi^R(u,x-x'),\qquad t>0.
\end{equation}
For the point source $\mu_r(x)=\mu_r\delta(x)$ and $t>|x|$, the time integral stays strictly inside the light cone, where Eq.~\eqref{eq:dirac-retarded-interior} can be integrated as an ordinary function.
At fixed $x$ and $T(t-|x|)\gg1$, using $F(t\pm x)\simeq2\pi T e^{-\pi T(t\pm x)}$ in Eq.~\eqref{eq:dirac-retarded-interior} gives
\begin{equation}\label{eq:dirac-fixed-x-retarded-tail}
 \Pi^R(t,x)\simeq4b^2\sin(2\pi\Delta)(2\pi T)^{4\Delta}
 \cosh(2\pi Tx)e^{-4\pi\Delta Tt}.
\end{equation}
Integrating this expression over time as in Eq.~\eqref{eq:local_step} gives
\begin{equation}\label{eq:dirac-point-source-fixed-x-tail}
 \delta\rho_r(x,t)
 \simeq 2\mu_rP_\Delta\cosh(2\pi Tx)e^{-4\pi\Delta Tt},
 \qquad P_\Delta=\frac{b^2\sin(2\pi\Delta)}{\Delta}(2\pi T)^{4\Delta-1}.
\end{equation}
Consequently, the late-time decay at a fixed position has rate $4\pi\Delta T$, with a position-dependent amplitude.

We now follow either ballistic front at a fixed distance $s=t-|x|>0$ behind the lightcone.
For $T(t-s)\gg1$ and $s$ beyond the microscopic cutoff, Eq.~\eqref{eq:dirac-retarded-interior} reduces to
\begin{equation}\label{eq:dirac-front-retarded-tail}
 \begin{aligned}
 \Pi^R\bigl(t,\pm(t-s)\bigr)
 &\simeq2b^2\sin(2\pi\Delta)F(2t-s)^{2\Delta-1}F(s)^{2\Delta+1}\\
 &\simeq2b^2\sin(2\pi\Delta)(2\pi T)^{2\Delta-1}
 e^{-\Gamma t}e^{\Gamma s/2}F(s)^{2\Delta+1}.
 \end{aligned}
\end{equation}
To obtain the imbalance along this trajectory, we evaluate the time integral in Eq.~\eqref{eq:local_step} at fixed $x$ and then set $x=\pm(t-s)$.
Changing variables to $u=t-s+y$ gives
\begin{equation}\label{eq:dirac-late-front-profile}
 \begin{aligned}
 \delta\rho_r\bigl(\pm(t-s),t\bigr)
 &\simeq 2\mu_r b^2\sin(2\pi\Delta)(2\pi T)^{2\Delta-1}e^{-\Gamma(t-s)}
 \int_s^\infty\dd y\,e^{-\pi T(2\Delta-1)y}
 \left[\frac{\pi T}{\sinh(\pi Ty)}\right]^{2\Delta+1}\\
 &=\mu_rP_\Delta\frac{e^{\Gamma s}}{\bigl(e^{2\pi Ts}-1\bigr)^{2\Delta}}e^{-\Gamma t}.
 \end{aligned}
\end{equation}
At fixed $s$, the prefactor is time independent and the imbalance decays at rate $\Gamma=2\pi T(2\Delta-1)$, the same as the spatially uniform imbalance in Eq.~\eqref{eq:dirac-field-removal-tail}.

\subsection{Endpoint limits of the Dirac conformal response}
\label{sec:dirac-endpoints}
The factor $\sin(2\pi\Delta)$ in Eq.~\eqref{eq:dirac-retarded-interior} vanishes at both ends of the interval $1/2<\Delta<1$.
In contrast to $d=0$, this does not imply that the entire positive-time response vanishes: distributions supported on the light cone must also be retained.

\paragraph{Lower endpoint:}
$\Delta \rightarrow 1/2$.
In this limit with $b\to b_0$ finite, the full Euclidean bubble in Eq.~\eqref{eq:dirac-thermal-density-bubble} reduces to
\[
 \Pi^E(\tau,x)=b_0^2\bigl[f_+^2+f_-^2\bigr].
\]
Analytically continuing to real time, the retarded bubble becomes:
\[
 \Pi^R(t,x)=-2\pi b_0^2\theta(t)\partial_t
 \bigl[\delta(x-t)+\delta(x+t)\bigr].
\]
For a localized source profile $\mu_r(x)=\mu_r\delta(x)$, inserting this into the spatial field-removal response, Eq.~\eqref{eq:local_step}, gives for $t>0$
\begin{equation}\label{eq:dirac-lower-endpoint-quench}
 \begin{aligned}
 \delta\rho_r(x,t)
 &=2\pi b_0^2\mu_r\bigl[\delta(x-t)+\delta(x+t)\bigr].
 \end{aligned}
\end{equation}
Thus $\Gamma=2\pi T(2\Delta-1)\to0$ corresponds to undamped ballistic propagation, consistent with the separate conservation of A and B charges in the free massless Dirac limit.

\paragraph{Upper endpoint:}
$\Delta\rightarrow1$.
This limit marks the boundary of the pure-power conformal solutions, corresponding to $M/(2N)\to1/4$.
Although the conformal decay exponent approaches $2\pi T$, its uniform-response residue $\mathcal A_\Delta$ vanishes for finite $b$, so this extrapolation does not determine the endpoint dynamics.
For $M/(2N)\leq1/4$, logarithmic corrections to the fermion propagator have been conjectured~\cite{KimAltmanCao2021}.
Determining charge relaxation in this regime requires the corresponding finite-temperature saddle and lies beyond the present analysis.

\section{Equilibrium linear-response numerics}\label{sec:numerics}
We use linear response to calculate the relaxation of a weak charge imbalance in complex SYK$_4$ and its Yukawa generalizations in $d=0,1$.
The relative-density susceptibility is given by a bubble of equilibrium real-time Green's functions, which we obtain by numerically solving the Schwinger--Dyson equations.
Figure~\ref{fig:decay-fits} shows representative decay curves and the exponential fits used to extract the relaxation rates.
All calculations are performed at charge neutrality.

\begin{figure}[t]
 \centering
 \includegraphics[width=\textwidth]{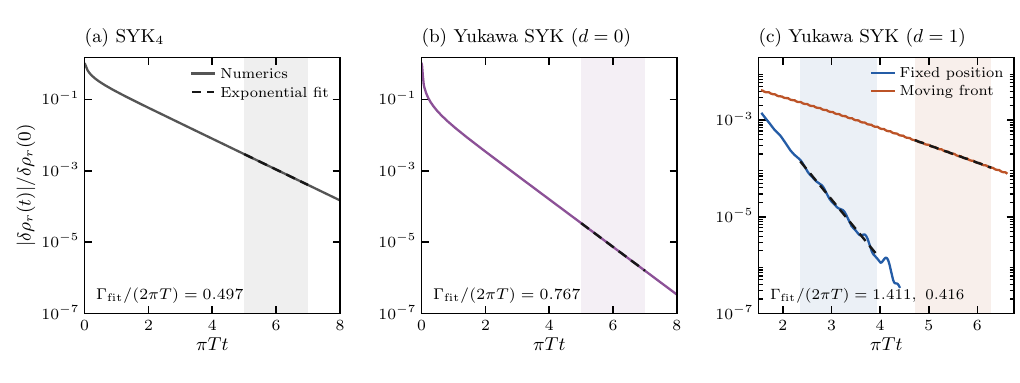}
 \caption{Relative-density relaxation and exponential fits for (a) complex SYK$_4$, (b) Yukawa SYK in $d=0$, and (c) Yukawa SYK in $d=1$.
 The temperatures are $T/J=0.00625$ in (a) and $T/g^2=0.00625$ in (b,c), with $\gamma=1$ for both Yukawa models and the parameters of Fig.~\ref{fig:main}.
 Solid curves show the magnitude of the linear-response imbalance after removing a weak source, normalized by its initial value in (a,b) and by $\delta\rho_r(0,0)$ for both curves in (c).
 Dashed lines are fits to $Ae^{-\Gamma_{\rm fit}t}$ over the shaded windows: $5<\pi Tt<7$ in (a,b), $0.75<Tt<1.25$ for the fixed-position curve in (c), and $1.5<Tt<2$ for the moving-front curve.
 In (c), both curves show the magnitude of a negative response behind the ballistic front: at $x=0$ (blue) and along $x=vt-6$ (orange), where $v\simeq0.5973$ is the measured velocity of the density peak.
 }
 \label{fig:decay-fits}
\end{figure}

\paragraph{Rate extraction.}
We extract $\Gamma_{\rm fit}$ by a linear least-squares fit of $\log|\delta\rho_r(t)|$ to $\log A-\Gamma_{\rm fit}t$, allowing both the amplitude $A$ and the rate to vary.
The fit has no additive offset, and each temperature is fitted independently.
To assess sensitivity to the fitting interval, we repeat the fit after translating the nominal window by $\pm20\%$ of its width: $\pm0.4$ in $\pi Tt$ for the two $d=0$ models and $\pm0.1$ in $Tt$ for the two $d=1$ trajectories.
In Fig.~\ref{fig:main}, the markers show the midpoint of the range of $\Gamma_{\rm fit}/(2\pi T)$ obtained from the three fitting windows, with error bars spanning that range.
These bars measure fitting-window sensitivity; discretization and source-profile effects are assessed separately below.

\subsection{Complex SYK$_4$}
For complex SYK$_4$, the equilibrium Schwinger--Dyson equations are
\begin{subequations}\label{eq:syk-numerical-saddle}
\begin{align}
 \mathcal G^R(E)&=[E-\Sigma^R(E)]^{-1},\\
 \Sigma^>(t)&=J^2[\mathcal G^>(t)]^2\mathcal G^<(-t),
 \qquad \Sigma^<(t)=J^2[\mathcal G^<(t)]^2\mathcal G^>(-t),\\
 \Sigma^R(t)&=\theta(t)[\Sigma^>(t)-\Sigma^<(t)],
\end{align}
\end{subequations}
where $\mathcal G^{>,<}$ are the greater and lesser Green's functions, and $\Sigma^{>,<}$ are the corresponding self-energies.
At equilibrium, the fluctuation--dissipation theorem relates these Green's functions to the spectral function $\mathcal A(E)=-\operatorname{Im}\mathcal G^R(E)/\pi$:
\begin{equation}\label{eq:syk-spectral-propagators}
 \mathcal G^>(t)=-\ii\int\dd E\,\mathcal A(E)[1-f(E)]e^{-\ii Et},\qquad
 \mathcal G^<(t)=\ii\int\dd E\,\mathcal A(E)f(E)e^{-\ii Et}.
\end{equation}
Here $f(E)=(e^{E/T}+1)^{-1}$ is the Fermi function.
Solving Eqs.~\eqref{eq:syk-numerical-saddle} and \eqref{eq:syk-spectral-propagators} self-consistently gives the greater and lesser Green's functions, which determine the retarded relative-density bubble:
\begin{equation}\label{eq:numerical-spectral-bubble}
 \Pi^R(t)=2\theta(t)\operatorname{Im}[\mathcal G^>(t)\mathcal G^<(t)^*].
\end{equation}
As discussed in the main text, this bubble gives the retarded relative-density correlator, $\chi_{rr}^R(t)=\Pi^R(t)/[N\epsilon(1-\epsilon)]$.
After removing the weak source at $t=0$, linear response gives the remaining density imbalance:
\begin{equation}\label{eq:numerical-spectral-response}
 \frac{\delta\rho_r(t)}{\mu_r}=\int_t^\infty\dd u\,\Pi^R(u).
\end{equation}

We take $J=1$ and evaluate ten equally spaced temperatures between $T=0.00625$ and $0.0125$.
We use a time step $\Delta t=0.025$ and $N_\omega=2^{18}$--$2^{19}$ frequency points, giving a frequency cutoff $\pi/\Delta t$ and a frequency spacing $\Delta\omega=2\pi/(N_\omega\Delta t)<0.08T$.
We integrate the response beyond $\pi Tt=64$, well past the fitting window $5<\pi Tt<7$.
At the lowest temperature, $T=0.00625$, doubling either the bandwidth or the Fourier time period changes the fitted rate by less than $1.4\times10^{-5}\%$, indicating good numerical convergence.

\subsection{Yukawa SYK in $d=0$}\label{sec:lowrank-numerics}
For Yukawa SYK in $d=0$, the equilibrium fermion and boson Green's functions obey the coupled Schwinger--Dyson equations
\begin{subequations}\label{eq:lowrank-numerical-saddle}
\begin{align}
 \mathcal G^R(\omega)&=[\omega-\Sigma^R(\omega)]^{-1},
 \qquad F^R(\omega)=-[1-\Pi_B^R(\omega)]^{-1},\\
 \Pi_B^R(t)&=\ii g^2\theta(t)
 [\mathcal G^>(t)\mathcal G^<(-t)-\mathcal G^<(t)\mathcal G^>(-t)],\\
 \Sigma^R(t)&=\ii\gamma g^2\theta(t)
 [F^>(t)\mathcal G^>(t)-F^<(t)\mathcal G^<(t)],
\end{align}
\end{subequations}
where $\gamma=M/N$, $F^{R,>,<}$ are the retarded, greater, and lesser boson Green's functions, and $\Pi_B^R$ is the retarded boson self-energy.

At equilibrium, the fermion Green's functions obey Eq.~\eqref{eq:syk-spectral-propagators}, while the fluctuation--dissipation theorem for the bosons gives
\begin{equation}\label{eq:boson-spectral-propagators}
 F^>(\omega)=-\ii[1+n_B(\omega)]B(\omega),\qquad
 F^<(\omega)=-\ii n_B(\omega)B(\omega),
\end{equation}
where $B(\omega)=-2\operatorname{Im}F^R(\omega)$ is the boson spectral function and $n_B(\omega)=(e^{\omega/T}-1)^{-1}$ is the Bose function.
Solving these equations self-consistently determines the equilibrium Green's functions.
The resulting fermion Green's functions give the relative-density bubble through Eq.~\eqref{eq:numerical-spectral-bubble}, whose time integral yields the imbalance after source removal through Eq.~\eqref{eq:numerical-spectral-response}.

Near zero frequency, the Bose function diverges as $n_B(\omega)\simeq T/\omega$, while the boson spectral function vanishes as $B(\omega)\simeq B'(0)\omega$.
The products $n_B(\omega)B(\omega)$ and $[1+n_B(\omega)]B(\omega)$ in Eq.~\eqref{eq:boson-spectral-propagators} therefore both approach the finite limit $TB'(0)$, which we use at the zero-frequency grid point.

We take $\gamma=1$ and $g=2$, with $g^2$ setting the energy scale in the chosen auxiliary-boson normalization.
We evaluate ten equally spaced temperatures between $T=0.025$ and $0.05$, covering the same normalized temperature range $0.00625\leq T/g^2\leq0.0125$ as SYK$_4$.
We use a time step $\Delta t=0.0125$ and $N_\omega=2^{17}$--$2^{18}$ frequency points, giving a frequency spacing $\Delta\omega<0.077T$.
We integrate the response beyond $\pi Tt=96$, well past the fitting window $5<\pi Tt<7$.
At the lowest temperature, $T=0.025$, doubling the bandwidth or halving the frequency spacing changes the fitted rate by less than $0.008\%$, indicating good numerical convergence.

To compare the fitted rates with the conformal prediction, we use the relation between the rank ratio and the fermion scaling dimension derived in Ref.~\cite{KimCaoAltman2020}:
\begin{equation}
 \gamma=\frac{(2\Delta-1)[\sec(2\pi\Delta)-1]}{4\Delta-1},
 \qquad \Delta\simeq0.384674\quad(\gamma=1).
\end{equation}
As the temperature decreases from $T=0.05$ to $0.025$, the nominal fitted rate $\Gamma_{\rm fit}/(2\pi T)$ increases from $0.7649$ to $0.7671$, indicating good agreement with the conformal field theory prediction $2\Delta\simeq0.7693$.

\subsection{Yukawa SYK in $d=1$}\label{sec:dirac-numerics}
For Yukawa SYK in $d=1$, we compute the spatially resolved imbalance to distinguish relaxation at fixed position from relaxation along the propagating front.
We work on a one-dimensional lattice with unit lattice spacing and nearest-neighbor hopping, giving the bare dispersion $\varepsilon_k=-t_1\cos k$.
At half filling, the chemical potential is zero and the Fermi points lie at $k=\pm\pi/2$.
The on-site interaction is built from zero-mean Gaussian Hermitian matrices $g^n$ that are identical at every site, preserving lattice translation symmetry.
Writing $c_{i,x}$ for the annihilation operator of flavor $i$ at site $x$ and taking normal ordering about half filling, the Hamiltonian is~\cite{KimAltmanCao2021}
\begin{equation}\label{eq:dirac-lattice-hamiltonian}
 \begin{aligned}
 H_{\rm lat}
 &=-\frac{t_1}{2}\sum_{x,i}\left(c_{i,x+1}^\dagger c_{i,x}+\mathrm{h.c.}\right)
 -\frac12\sum_{x,ij,kl}J_{ij,kl}:\!\left(c_{i,x}^\dagger c_{j,x}\right)\left(c_{k,x}^\dagger c_{l,x}\right)\!:\,,\\
 J_{ij,kl}&=\sum_{n=1}^{M}g_{ij}^n g_{kl}^n,
 \qquad \overline{g_{ij}^n g_{kl}^m}=\frac{g^2}{N^2}\delta_{il}\delta_{jk}\delta_{nm}.
 \end{aligned}
\end{equation}
Decoupling the on-site interaction through a Hubbard--Stratonovich transformation gives a Yukawa representation whose equilibrium Green's functions obey
\begin{subequations}\label{eq:dirac-numerical-saddle}
\begin{align}
 \mathcal G^R(k,\omega)&=[\omega-\varepsilon_k-\Sigma^R(k,\omega)]^{-1},
 \qquad F^R(q,\omega)=-[1-\Pi_B^R(q,\omega)]^{-1},\\
 \Pi_B^R(x,t)&=\ii g^2(-1)^x\theta(t)
 \left[\mathcal G^>(x,t)\mathcal G^<(-x,-t)-\mathcal G^<(x,t)\mathcal G^>(-x,-t)\right],\\
 \Sigma^R(x,t)&=2\ii\gamma g^2(-1)^x\theta(t)
 \left[F^>(x,t)\mathcal G^>(x,t)-F^<(x,t)\mathcal G^<(x,t)\right],
\end{align}
\end{subequations}
where $\mathcal G^R(k,\omega)$ is the scalar lattice-fermion Green's function and $\gamma=M/(2N)$.
We use staggered HS-boson fields so that the mode connecting the two Fermi points lies at zero boson momentum; this convention produces the factors $(-1)^x$ in the self-energies.
At equilibrium, Eqs.~\eqref{eq:syk-spectral-propagators} and \eqref{eq:boson-spectral-propagators} apply at each momentum and determine the greater and lesser Green's functions from the fermion and boson spectral functions.

Solving these equations self-consistently gives the fermion spectral function $\mathcal A(k,E)=-\operatorname{Im}\mathcal G^R(k,E)/\pi$.
We Fourier transform the greater and lesser Green's functions to real space and construct the relative-density bubble:
\begin{equation}\label{eq:numerical-spatial-bubble}
 \Pi^R(x,t)=2\theta(t)\operatorname{Im}[\mathcal G^>(x,t)\mathcal G^<(x,t)^*].
\end{equation}
We prepare the imbalance with a weak localized source $\mu_r(x)=\mu_r w(x)$, where $w(x)\propto e^{-x^2/(2\sigma^2)}$ has width $\sigma=3$ lattice sites and is normalized by $\sum_xw(x)=1$.
This smooth profile suppresses the alternating lattice-density component.
After removing the source at $t=0$, linear response gives
\begin{equation}\label{eq:numerical-spatial-response}
 \frac{\delta\rho_r(x,t)}{\mu_r}=\sum_y w(y)\int_t^\infty\dd u\,\Pi^R(x-y,u).
\end{equation}
We evaluate the time integral at fixed position before sampling the resulting response along a moving trajectory.

We take $t_1=1$ (unit bare Fermi velocity), $\gamma=1$, and $g=2$, and evaluate ten equally spaced temperatures between $T=0.025$ and $0.05$, covering the same normalized range $0.00625\leq T/g^2\leq0.0125$ as Yukawa SYK in $d=0$.
We use $N_x=2^8$ momentum points, a time step $\Delta t=0.0625$, and $N_\omega=2^{13}$--$2^{14}$ frequency points, giving a frequency spacing $\Delta\omega<0.246T$.
We integrate the response beyond $\pi Tt=30$, well past both fitting windows specified below.

To extract the two decay rates, we first determine the front velocity by fitting the position of the outward-moving density peak linearly in time over $0.5<Tt<2$.
This gives $v\simeq0.5973$ at $T=0.025$.
We fit the response magnitude at a fixed position $x=0$ over $0.75<Tt<1.25$ and along the moving trajectory $x=vt-6$ over $1.5<Tt<2$.
The source width and fitting windows are the same at every temperature, while $v$ and the decay rates are determined separately for each temperature.

At $T=0.025$, increasing $N_x$ from $2^7$ to $2^8$ changes both fitted rates by less than $0.004\%$.
At $N_x=2^7$, doubling the frequency cutoff at fixed frequency spacing changes both rates by less than $0.1\%$, with the peak velocity redetermined for each cutoff.
Increasing the source width from three to four sites changes the rates by less than $2.5\%$ at this temperature.

The negative prefactor $P_\Delta$ in Eqs.~\eqref{eq:quench_d1} and \eqref{eq:quench_d1_front} implies a negative response $\delta\rho_r/\mu_r$ behind the ballistic front, consistent with both low-temperature numerical trajectories in Fig.~\ref{fig:decay-fits}(c).
Both trajectories remain negative throughout the nominal and shifted fitting windows at all ten temperatures.
The full rate spread is below $8\%$ at fixed position and $3.1\%$ along the moving trajectory.

At the lowest temperature, $T=0.025$, the nominal fitted rates are $\Gamma_{\rm fit}/(2\pi T)=1.4107$ at fixed position and $0.4158$ along the moving front, close to the respective conformal predictions $1.4$ and $0.4$.
\end{widetext}

\end{document}